\pdfoutput=1
\documentclass[aps,prd,reprint,superscriptaddress,showpacs]{revtex4-1}

\usepackage{graphicx}
\usepackage{dcolumn}
\usepackage{color}
\usepackage{xspace}
\usepackage{multirow}

\newcommand{\ppbar}{$p\bar{p}$}
\newcommand{\roots}{$\sqrt{s}$}
\newcommand{\fb}{fb$^{-1}$}

\newcommand{\gevc}{GeV/$c$}
\newcommand{\gevcc}{GeV/$c^2$}

\newcommand{\mgg}{$m_{\gamma\gamma}$}
\newcommand{\ptgg}{$p_T^{\gamma\gamma}$}
\newcommand{\hf}{$h_f$}
\newcommand{\Mh}{$m_H$}
\newcommand{\Mhf}{$m_{h_f}$}
\newcommand{\Hgg}{$H\rightarrow\gamma \gamma$}
\newcommand{\Hfgg}{$h_f\rightarrow\gamma \gamma$}
\newcommand{\BRHgg}{$\mathcal{B}(H\rightarrow\gamma\gamma)$}
\newcommand{\BRHfgg}{$\mathcal{B}(h_f\rightarrow\gamma\gamma)$}
\newcommand{\py}{\textsc{pythia}}
\newcommand{\mc}{\multicolumn}
\newcommand{\CCc}{\mbox{C$^\prime$C}\xspace}
\newcommand{\CPc}{\mbox{C$^\prime$P}\xspace}
\newcommand{\gjjj}{$\gamma j$ + {\it jj}}

\begin{document}

\title{Search for a Higgs boson in the diphoton final state using the full CDF data set \\ from \ppbar\ collisions at \roots~=~1.96~TeV}

\affiliation{Institute of Physics, Academia Sinica, Taipei, Taiwan 11529, Republic of China}
\affiliation{Argonne National Laboratory, Argonne, Illinois 60439, USA}
\affiliation{University of Athens, 157 71 Athens, Greece}
\affiliation{Institut de Fisica d'Altes Energies, ICREA, Universitat Autonoma de Barcelona, E-08193, Bellaterra (Barcelona), Spain}
\affiliation{Baylor University, Waco, Texas 76798, USA}
\affiliation{Istituto Nazionale di Fisica Nucleare Bologna, $^{ee}$University of Bologna, I-40127 Bologna, Italy}
\affiliation{University of California, Davis, Davis, California 95616, USA}
\affiliation{University of California, Los Angeles, Los Angeles, California 90024, USA}
\affiliation{Instituto de Fisica de Cantabria, CSIC-University of Cantabria, 39005 Santander, Spain}
\affiliation{Carnegie Mellon University, Pittsburgh, Pennsylvania 15213, USA}
\affiliation{Enrico Fermi Institute, University of Chicago, Chicago, Illinois 60637, USA}
\affiliation{Comenius University, 842 48 Bratislava, Slovakia; Institute of Experimental Physics, 040 01 Kosice, Slovakia}
\affiliation{Joint Institute for Nuclear Research, RU-141980 Dubna, Russia}
\affiliation{Duke University, Durham, North Carolina 27708, USA}
\affiliation{Fermi National Accelerator Laboratory, Batavia, Illinois 60510, USA}
\affiliation{University of Florida, Gainesville, Florida 32611, USA}
\affiliation{Laboratori Nazionali di Frascati, Istituto Nazionale di Fisica Nucleare, I-00044 Frascati, Italy}
\affiliation{University of Geneva, CH-1211 Geneva 4, Switzerland}
\affiliation{Glasgow University, Glasgow G12 8QQ, United Kingdom}
\affiliation{Harvard University, Cambridge, Massachusetts 02138, USA}
\affiliation{Division of High Energy Physics, Department of Physics, University of Helsinki and Helsinki Institute of Physics, FIN-00014, Helsinki, Finland}
\affiliation{University of Illinois, Urbana, Illinois 61801, USA}
\affiliation{The Johns Hopkins University, Baltimore, Maryland 21218, USA}
\affiliation{Institut f\"{u}r Experimentelle Kernphysik, Karlsruhe Institute of Technology, D-76131 Karlsruhe, Germany}
\affiliation{Center for High Energy Physics: Kyungpook National University, Daegu 702-701, Korea; Seoul National University, Seoul 151-742, Korea; Sungkyunkwan University, Suwon 440-746, Korea; Korea Institute of Science and Technology Information, Daejeon 305-806, Korea; Chonnam National University, Gwangju 500-757, Korea; Chonbuk National University, Jeonju 561-756, Korea}
\affiliation{Ernest Orlando Lawrence Berkeley National Laboratory, Berkeley, California 94720, USA}
\affiliation{University of Liverpool, Liverpool L69 7ZE, United Kingdom}
\affiliation{University College London, London WC1E 6BT, United Kingdom}
\affiliation{Centro de Investigaciones Energeticas Medioambientales y Tecnologicas, E-28040 Madrid, Spain}
\affiliation{Massachusetts Institute of Technology, Cambridge, Massachusetts 02139, USA}
\affiliation{Institute of Particle Physics: McGill University, Montr\'{e}al, Qu\'{e}bec, Canada H3A~2T8; Simon Fraser University, Burnaby, British Columbia, Canada V5A~1S6; University of Toronto, Toronto, Ontario, Canada M5S~1A7; and TRIUMF, Vancouver, British Columbia, Canada V6T~2A3}
\affiliation{University of Michigan, Ann Arbor, Michigan 48109, USA}
\affiliation{Michigan State University, East Lansing, Michigan 48824, USA}
\affiliation{Institution for Theoretical and Experimental Physics, ITEP, Moscow 117259, Russia}
\affiliation{University of New Mexico, Albuquerque, New Mexico 87131, USA}
\affiliation{The Ohio State University, Columbus, Ohio 43210, USA}
\affiliation{Okayama University, Okayama 700-8530, Japan}
\affiliation{Osaka City University, Osaka 588, Japan}
\affiliation{University of Oxford, Oxford OX1 3RH, United Kingdom}
\affiliation{Istituto Nazionale di Fisica Nucleare, Sezione di Padova-Trento, $^{ff}$University of Padova, I-35131 Padova, Italy}
\affiliation{University of Pennsylvania, Philadelphia, Pennsylvania 19104, USA}
\affiliation{Istituto Nazionale di Fisica Nucleare Pisa, $^{gg}$University of Pisa, $^{hh}$University of Siena and $^{ii}$Scuola Normale Superiore, I-56127 Pisa, Italy}
\affiliation{University of Pittsburgh, Pittsburgh, Pennsylvania 15260, USA}
\affiliation{Purdue University, West Lafayette, Indiana 47907, USA}
\affiliation{University of Rochester, Rochester, New York 14627, USA}
\affiliation{The Rockefeller University, New York, New York 10065, USA}
\affiliation{Istituto Nazionale di Fisica Nucleare, Sezione di Roma 1, $^{jj}$Sapienza Universit\`{a} di Roma, I-00185 Roma, Italy}
\affiliation{Rutgers University, Piscataway, New Jersey 08855, USA}
\affiliation{Texas A\&M University, College Station, Texas 77843, USA}
\affiliation{Istituto Nazionale di Fisica Nucleare Trieste/Udine, I-34100 Trieste, $^{kk}$University of Udine, I-33100 Udine, Italy}
\affiliation{University of Tsukuba, Tsukuba, Ibaraki 305, Japan}
\affiliation{Tufts University, Medford, Massachusetts 02155, USA}
\affiliation{University of Virginia, Charlottesville, Virginia 22906, USA}
\affiliation{Waseda University, Tokyo 169, Japan}
\affiliation{Wayne State University, Detroit, Michigan 48201, USA}
\affiliation{University of Wisconsin, Madison, Wisconsin 53706, USA}
\affiliation{Yale University, New Haven, Connecticut 06520, USA}

\author{T.~Aaltonen}
\affiliation{Division of High Energy Physics, Department of Physics, University of Helsinki and Helsinki Institute of Physics, FIN-00014, Helsinki, Finland}
\author{B.~\'{A}lvarez~Gonz\'{a}lez$^z$}
\affiliation{Instituto de Fisica de Cantabria, CSIC-University of Cantabria, 39005 Santander, Spain}
\author{S.~Amerio}
\affiliation{Istituto Nazionale di Fisica Nucleare, Sezione di Padova-Trento, $^{ff}$University of Padova, I-35131 Padova, Italy}
\author{D.~Amidei}
\affiliation{University of Michigan, Ann Arbor, Michigan 48109, USA}
\author{A.~Anastassov$^x$}
\affiliation{Fermi National Accelerator Laboratory, Batavia, Illinois 60510, USA}
\author{A.~Annovi}
\affiliation{Laboratori Nazionali di Frascati, Istituto Nazionale di Fisica Nucleare, I-00044 Frascati, Italy}
\author{J.~Antos}
\affiliation{Comenius University, 842 48 Bratislava, Slovakia; Institute of Experimental Physics, 040 01 Kosice, Slovakia}
\author{G.~Apollinari}
\affiliation{Fermi National Accelerator Laboratory, Batavia, Illinois 60510, USA}
\author{J.A.~Appel}
\affiliation{Fermi National Accelerator Laboratory, Batavia, Illinois 60510, USA}
\author{T.~Arisawa}
\affiliation{Waseda University, Tokyo 169, Japan}
% ...

\collaboration{CDF Collaboration}
\thanks{With visitors from
$^a$Istituto Nazionale di Fisica Nucleare, Sezione di Cagliari, 09042 Monserrato (Cagliari), Italy,
$^b$University of CA Irvine, Irvine, CA 92697, USA,
$^c$University of CA Santa Barbara, Santa Barbara, CA 93106, USA,
$^d$University of CA Santa Cruz, Santa Cruz, CA 95064, USA,
$^e$Institute of Physics, Academy of Sciences of the Czech Republic, Czech Republic,
$^f$CERN, CH-1211 Geneva, Switzerland,
$^g$Cornell University, Ithaca, NY 14853, USA,
$^h$University of Cyprus, Nicosia CY-1678, Cyprus,
$^i$Office of Science, U.S. Department of Energy, Washington, DC 20585, USA,
$^j$University College Dublin, Dublin 4, Ireland,
$^k$ETH, 8092 Zurich, Switzerland,
$^l$University of Fukui, Fukui City, Fukui Prefecture, Japan 910-0017,
$^m$Universidad Iberoamericana, Mexico D.F., Mexico,
$^n$University of Iowa, Iowa City, IA 52242, USA,
$^o$Kinki University, Higashi-Osaka City, Japan 577-8502,
$^p$Kansas State University, Manhattan, KS 66506, USA,
$^q$Ewha Womans University, Seoul, 120-750, Korea,
$^r$University of Manchester, Manchester M13 9PL, United Kingdom,
$^s$Queen Mary, University of London, London, E1 4NS, United Kingdom,
$^t$University of Melbourne, Victoria 3010, Australia,
$^u$Muons, Inc., Batavia, IL 60510, USA,
$^v$Nagasaki Institute of Applied Science, Nagasaki, Japan,
$^w$National Research Nuclear University, Moscow, Russia,
$^x$Northwestern University, Evanston, IL 60208, USA,
$^y$University of Notre Dame, Notre Dame, IN 46556, USA,
$^z$Universidad de Oviedo, E-33007 Oviedo, Spain,
$^{aa}$CNRS-IN2P3, Paris, F-75205 France,
$^{bb}$Texas Tech University, Lubbock, TX 79609, USA,
$^{cc}$Universidad Tecnica Federico Santa Maria, 110v Valparaiso, Chile,
$^{dd}$Yarmouk University, Irbid 211-63, Jordan
}
\noaffiliation

\date{\today}

\begin{abstract}
A search for a narrow Higgs boson resonance in the diphoton 
mass spectrum is  presented based on data corresponding to 
10~\fb\ of integrated luminosity from proton-antiproton collisions 
at \roots~= 1.96 TeV collected by the CDF experiment. In 
addition to searching for a resonance in the diphoton mass 
spectrum, we employ a multivariate discriminant technique for 
the first time in this channel at CDF. No evidence of signal is 
observed, and upper limits are set on the cross section times 
branching ratio of the resonant state as a function of Higgs 
boson mass.  The limits  are interpreted in the context of the 
standard model with an expected (observed) limit on the cross 
section times branching ratio of 9.9 (17.0) times the standard 
model prediction at the 95\% credibility level for a Higgs boson 
mass of 125~\gevcc. Moreover, a Higgs boson with suppressed 
couplings to fermions is excluded for masses below 114~\gevcc\ 
at the 95\% credibility level.
\end{abstract}


\maketitle

\section{Introduction}
\label{sec:introduction}

The standard model (SM) of particle physics has proven to be a 
robust theory that accurately describes the properties of elementary 
particles and the forces of interaction between them. However, the 
origin of mass has remained an unsolved mystery for decades. The 
SM suggests that particles acquire mass due to interactions with the 
Higgs field via spontaneous symmetry breaking~\cite{Higgs1964,
*Guralnik:1964eu,*Englert:1964et}.  Direct searches at the 
Large Electron-Positron Collider (LEP)~\cite{lephiggs}, combined 
with recent search results from the Tevatron~\cite{Tevhiggs} and the 
Large Hadron Collider (LHC)~\cite{atlasnewesthiggs,cmshiggs}, 
exclude all potential SM Higgs boson masses outside the ranges 
116.6--119.4~\gevcc\ and 122.1--127~\gevcc.

In the SM, the branching ratio for a Higgs boson decaying into a 
photon pair \BRHgg\ is maximal for Higgs boson masses between 
about 110 and 140~\gevcc. This is a mass range that is most useful 
for Higgs boson searches at the Fermilab Tevatron~\cite{Tevhiggs} 
and is favored by indirect constraints from electroweak 
observables~\cite{EwkLimits2010}. The SM \Hgg\ branching ratio 
peaks at a value of about 0.23\% for a Higgs boson mass \Mh~= 
125~\gevcc~\cite{Djouadi:1997yw}. This is a very small branching 
ratio; however, the distinctive signal that photons produce in the 
detector makes \Hgg\ an appealing search mode. Compared to the 
dominant decay modes involving $b$ quarks, a larger fraction of 
\Hgg\ events can be identified and the diphoton invariant mass of 
these events would cluster in a narrower range, thus providing 
a better discriminator against the smoothly distributed background. 
There are also theories beyond the standard model that predict a 
suppressed coupling of a Higgs boson to fermions. In these 
``fermiophobic'' Higgs boson models, the diphoton decay can be 
greatly enhanced~\cite{Rosca:2002me, *Akeroyd:1995hg, 
*Portugal, *Mrenna, *Haber:1978jt, *Gunion:1989ci, *Barger:1992ty, 
*Basdevant:1992nb, *PhysRevD.49.1354, *Diaz:1994pk, 
*bruchersantos, *Matchev1}.

The Collider Detector at Fermilab (CDF) and D0 experiments at the Tevatron 
have searched for both a SM Higgs boson, $H$, and a fermiophobic Higgs 
boson, \hf, decaying to two photons~\cite{Affolder:2001hx, Abbott:1998vv, 
hfD0, Aaltonen:2009ga}. The CDF and D0 experiments recently set 95\% 
credibility level (C.L.) upper limits on the cross section times branching ratio 
$\sigma$~$\times$ \BRHgg\ relative to the SM prediction and on \BRHfgg\ 
using data corresponding to an integrated luminosity $\cal{L}$ of 7.0~\fb~
\cite{cdfHgg2012} and 8.2~\fb~\cite{d0Higgs2011}, respectively. The \hf\ 
result sets a lower limit on \Mhf\ of   114~\gevcc\ and 112.9~\gevcc, 
respectively. These results surpassed for the first time the 109.7~\gevcc\ 
mass limit obtained from combined searches at the LEP collider at CERN~
\cite{Rosca:2002me}. 

Recently, the ATLAS and CMS experiments at the LHC at CERN have 
searched for a SM Higgs boson decaying to two photons using $\cal{L}$ = 
4.9~\fb~\cite{atlasHgg2012} and 4.8~\fb~\cite{cmsHgg2012}, respectively.   
In the low mass range, rates corresponding to less than twice the SM cross 
section are excluded at 95\% C.L.  An excess of nearly $2\sigma$ is present 
in both the CMS and ATLAS results, which could be consistent with a SM 
Higgs boson with a mass near 125~\gevcc.  

In this Letter, we present a search for a Higgs boson decaying to two photons  
using the final CDF diphoton data set, corresponding to an integrated 
luminosity of 10~\fb. This analysis searches the diphoton mass distribution for 
a narrow resonance that could reveal the presence of a SM or fermiophobic 
Higgs boson, updating the previous CDF result~\cite{cdfHgg2012} with more 
than 40\% additional integrated luminosity. We furthermore implement a new 
multivariate technique for events that contain two central photons, using both 
diphoton and jet kinematic variables to improve the sensitivity for identifying a 
Higgs boson signal from the diphoton backgrounds.   

\section{Higgs boson signal model}
\label{sec:Signal}

For the SM search, we consider the three most likely production mechanisms at the 
Tevatron: gluon fusion (GF); associated production (VH), where a Higgs boson is 
produced in association with a $W$ or $Z$ boson; and vector boson fusion (VBF), 
where a Higgs boson is produced alongside two quark jets. As an example, the SM 
cross sections for \Mh~= 125~\gevcc\ are 949.3~fb~\cite{Anastasiou:2008tj, 
*deFlorian:2009hc}, 208.0~fb~\cite{Baglio:2010um, *Brein:2003wg, 
*Ciccolini:2003jy}, and 65.3~fb~\cite{Bolzoni:2010xr, *Ciccolini:2007jr, *Ciccolini:2007ec}, 
respectively. In the fermiophobic search, we consider a benchmark model in which a 
Higgs boson does not couple to fermions, yet retains its SM couplings to bosons~
\cite{Akeroyd:1995hg}.  In this model, the GF process is suppressed and 
fermiophobic Higgs boson production is dominated by VH and VBF. With $\cal{L}$ = 
10~\fb, about 28 (43) \Hgg\ (\Hfgg) events are predicted to be produced for \Mh~= 
125~\gevcc.

Only about 25\% of these events would produce photons that are absorbed in 
well-instrumented regions of the CDF detector and pass the full diphoton selection  
discussed in Section~\ref{sec:selection}~\cite{cdfHgg2012}. This fraction, along 
with the predicted distributions of kinematic variables, is obtained from a 
simulation of Higgs boson decays into diphotons.  For each Higgs boson mass 
hypothesis tested in the range 100--150~\gevcc, in 5~\gevcc\ steps, signal 
samples are developed from the \py\ 6.2~\cite{PYTHIA} Monte Carlo (MC) event 
generator and a parametrized response of the CDF II detector~\cite
{GEANT3,GFLASH}. All \py\ samples were made with CTEQ5L~\cite{CTEQ5L} 
parton distribution functions, where the \py\ underlying event model is tuned to 
CDF jet data~\cite{TUNEA}.  Each signal sample is corrected for multiple 
interactions and differences between the identification of photons in the simulation 
and the data~\cite{cdfHgg2012}.  The GF signal is furthermore corrected based on 
a higher-order theoretical prediction of the transverse momentum 
distribution~\cite{Bozzi:2003jy,*Bozzi:2005wk,*deFlorian:2011xf}.

\section{Detector and event selection}
\label{sec:selection}

We use the CDF II detector~\cite{Acosta:2004yw} to identify photon candidate 
events produced in \ppbar\ collisions at \roots~= 1.96 TeV. The silicon vertex 
tracker~\cite{Sill:2000zz} and the central outer tracker~\cite{Affolder:2003ep}, 
contained within a 1.4~T axial magnetic field, measure the trajectories of charged 
particles and determine their momenta. Particles that pass through the outer 
tracker reach the electromagnetic (EM) and hadronic calorimeters~\cite{ccal, wha, 
Albrow:2001jw}, which are divided into two regions: central ($|\eta|<1.1$) and 
forward or ``plug'' ($1.1<|\eta|<3.6$). The EM calorimeters contain fine-grained 
shower maximum detectors~\cite{Apollinari:1998bg}, which measure the shower 
shape and centroid position in the plane transverse to the direction of the shower 
development.

The event selection is the same as in the previous \Hgg\ search~\cite{cdfHgg2012}. 
Events with two photon candidates are selected and the data are divided into four 
independent categories according to the position and type of the photons. In 
central-central (CC) events, both photon candidates are detected within the fiducial 
region of the central EM calorimeter ($|\eta|< 1.05$);  in central-plug (CP) events, 
one photon candidate is detected in this region and the other is in the fiducial 
region of the plug calorimeter ($1.2<|\eta|<2.8$); in central-central events with a 
conversion (C$^\prime$C), both photon candidates are in the central region, but 
one photon converts and is reconstructed from its $e^+e^-$ decay products; in 
central-plug events with a conversion (C$^\prime$P), there is one central 
conversion candidate together with a plug photon candidate.

In order to improve sensitivity for the fermiophobic Higgs boson search, the event 
selection is extended by taking advantage of the final-state features present in the 
VH and VBF processes. Because the Higgs boson from these processes will be 
produced in association with a $W$ or $Z$ boson, or with two jets, the transverse 
momentum of the diphoton system \ptgg\ is generally higher relative to the diphoton 
backgrounds. A requirement of \ptgg~$>75$~\gevc\ isolates a region of high \hf\ 
sensitivity, retaining roughly 30\% of the signal while removing 99.5\% of the 
background~\cite{Aaltonen:2009ga}. Two lower-\ptgg\ regions, \ptgg~$<35$~\gevc\ 
and $35$~\gevc~$<$ \ptgg~$<75$~\gevc, are additionally included and provide 
about 15\% more sensitivity to the \hf\ signal. With four diphoton categories (CC, 
CP, C$^\prime$C, and C$^\prime$P) and three \ptgg\ regions, twelve independent 
channels are included for the fermiophobic Higgs boson search. 

\begin{figure}
\includegraphics[width=8.66cm]{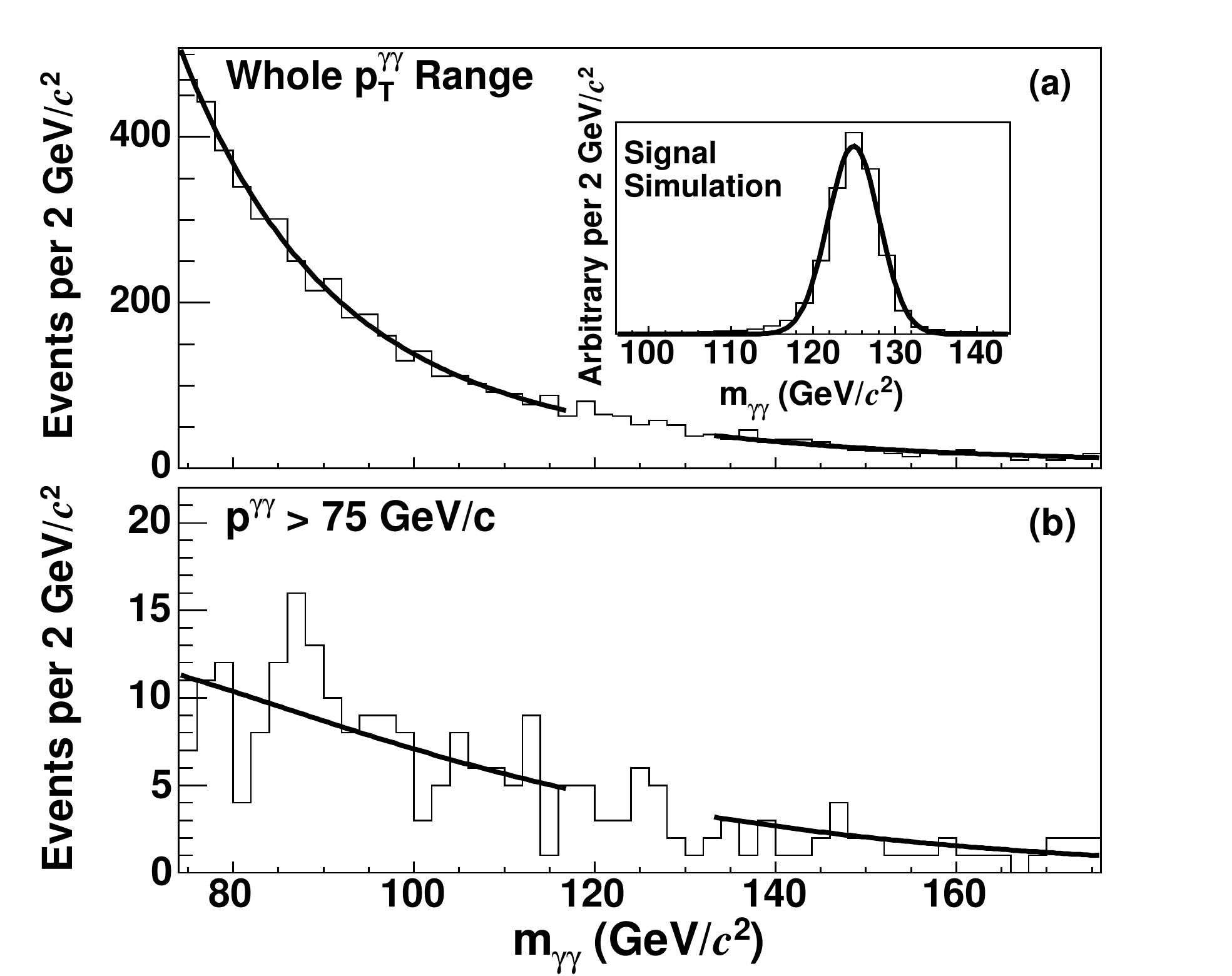}
\caption{ The invariant mass distribution of CC photon pairs in the data is shown for 
(a) the entire \ptgg\ region used in the SM Higgs boson diphoton resonance search 
and (b) the highest-\ptgg\ region (the most sensitive region) used in the \hf\ diphoton 
resonance search. Each distribution shows a fit to the data for the hypothesis of \Mh\ 
= 125~\gevcc, for which the signal region centered at 125~\gevcc\  is excluded from 
the fit.  The expected shape of the signal from simulation is shown in the inset of (a).
\label{fig:Mgg}}
\end{figure}

\section{Diphoton resonance search}
\label{sec:Mgg}

The decay of a Higgs boson into a diphoton pair would appear as a very narrow 
peak in the distribution of the invariant mass \mgg\ of the two photons. The diphoton 
mass resolution as determined from simulation is better than 3\%\ for the Higgs 
boson mass region studied here and is limited by the energy resolution of the 
electromagnetic calorimeters \cite{phores} and the ability to identify the primary 
interaction vertex~\cite{cdfHgg2012}. The diphoton invariant mass distribution for 
the most sensitive search  category in the SM and fermiophobic scenarios is 
provided in Fig.~\ref{fig:Mgg}, with an inset showing the signal shape expected from 
simulation. In each diphoton category, we perform a search of the \mgg\ spectrum  
for signs of a resonance.

For this search, the total diphoton background is modeled from a fit to the binned 
diphoton mass spectrum of the data using a log-likelihood ($\log {\cal L}$) method, 
as described in \cite{cdfHgg2012}. The fit is performed independently for each 
diphoton category and includes only the sideband region for each \Mh\ hypothesis, 
which is the control region excluding a mass window centered on the Higgs boson 
mass being tested. The full width of the mass window is chosen to be approximately 
$\pm2$ standard deviations of the expected Higgs boson mass resolution, which 
amounts to 12~\gevcc, 16~\gevcc, and 20~\gevcc\ for mass hypotheses of  
100--115~GeV/c$^2$, 120--135~GeV/c$^2$, and 140--150~GeV/c$^2$, 
respectively.  The fit for the CC category for \Mh\ = 125~\gevcc\ is shown in 
Fig.~\ref{fig:Mgg}. 

\section{Multivariate discriminator}
\label{sec:mva}

The diphoton mass distribution is the most powerful variable for separating a Higgs 
boson signal from the diphoton backgrounds. However, other information is 
available that can be used to further distinguish this signal. We improve the most 
sensitive search category (CC) by using a ``Multi-Layer Perceptron'' neural 
network (NN)~\cite{tmva}, which combines the information of several well-modeled 
kinematic variables into a single discriminator, optimized to separate signal and 
background events.  Four diphoton kinematic variables are included: \mgg, \ptgg, the 
difference between the azimuthal angles of the two photons, and the cosine of the 
photon emission angle relative to the colliding hadrons in the diphoton rest frame 
(the Collins-Soper angle)~\cite{Aaltonen:2011vk}. For events with jets, we also 
include four variables related to the jet activity, which are particularly useful for 
identifying VBF and VH signal events.  These variables are the number of jets in 
the event, the sum of the jet transverse energies, and the event sphericity and 
aplanarity~\cite{sphap}. Jets are reconstructed from tower clusters in the hadronic 
calorimeter within a cone of radius 0.4 in the $\eta-\phi$ plane~\cite{Bhatti:2005ai}. 
Each jet is required to have $\vert\eta\vert<2$ and a transverse energy 
$E_T>20$~GeV, where the energy is corrected for calorimeter response, multiple 
interactions, and absolute energy scale. 

In order to optimize the performance of the method, we divide the CC category into 
two independent subsamples of events: the CC0 category for events with no jets and 
the CCJ category for events with at least one jet. The CC0 category uses a network 
trained with only the four diphoton variables; the CCJ category uses a network 
trained with the four diphoton and four jet variables.

The sideband fit used in the diphoton resonance search provides an estimate of the total 
background prediction in each signal mass window; however, the multivariate analysis 
requires a more detailed background model.   Specifically, we divide the background into 
its distinct components in order to best model all input variables used by the discriminant, 
which is also sensitive to correlations.   There are two main background components in 
the CC data sample: a prompt diphoton ($\gamma\gamma$) background produced from 
the hard parton scattering or from hard photon bremsstrahlung from energetic quarks, 
and a background comprised of $\gamma$-jet and jet-jet events (\gjjj\/) in which the jets 
are misidentified as photons~\cite{phojet}. To model the shape of kinematic variables in 
the $\gamma\gamma$ background, we use a \py\ MC sample developed and studied in a 
measurement of the diphoton cross section~\cite{Aaltonen:2011vk}. To model the 
variable shapes in the \gjjj\ background, we obtain a data sample enriched in 
misidentified photons by selecting events for which one or both photon candidates fail the 
NN photon ID requirement~\cite{cdfHgg2012}.

\begin{figure}
\includegraphics[width=8.0cm]{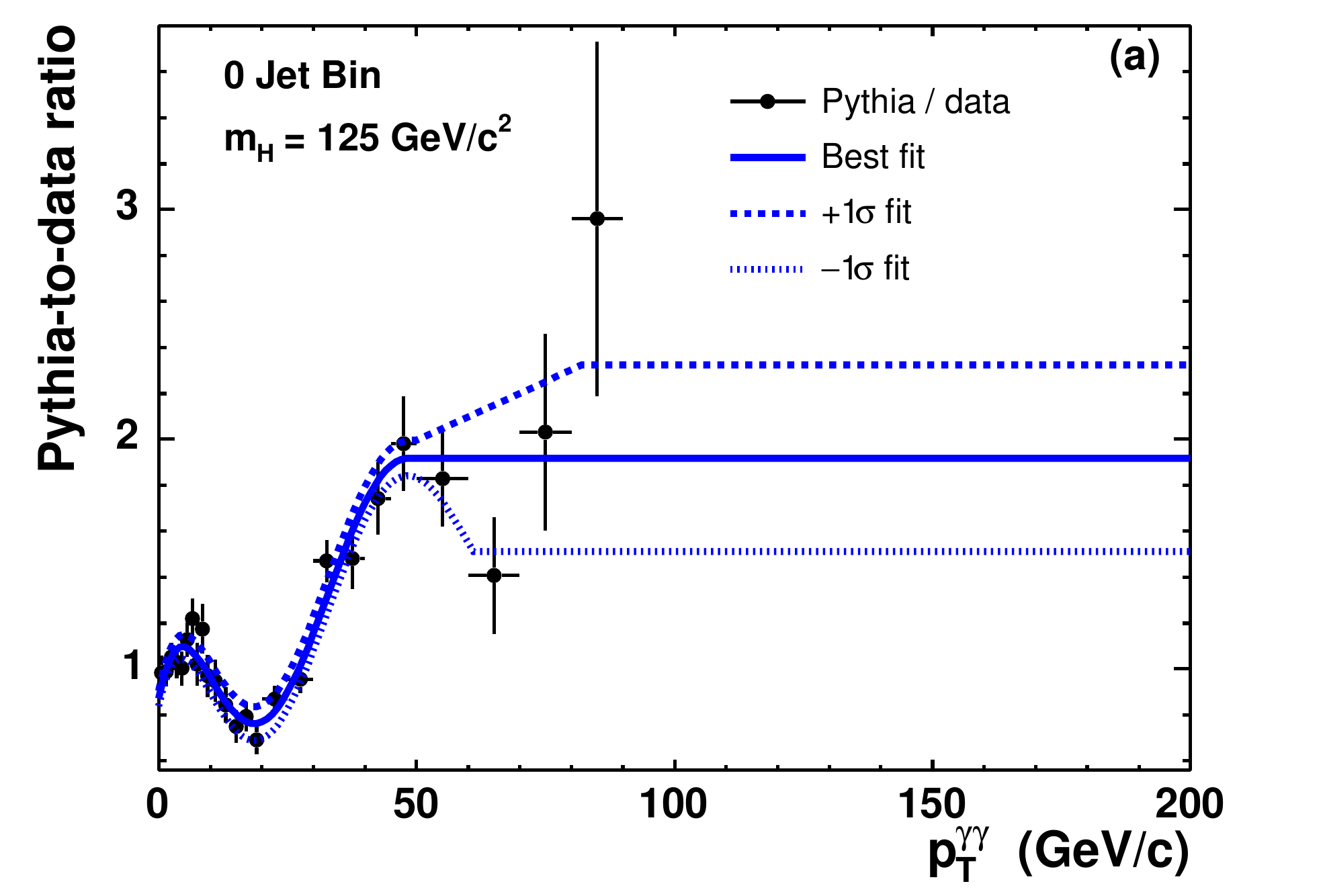}
\includegraphics[width=8.0cm]{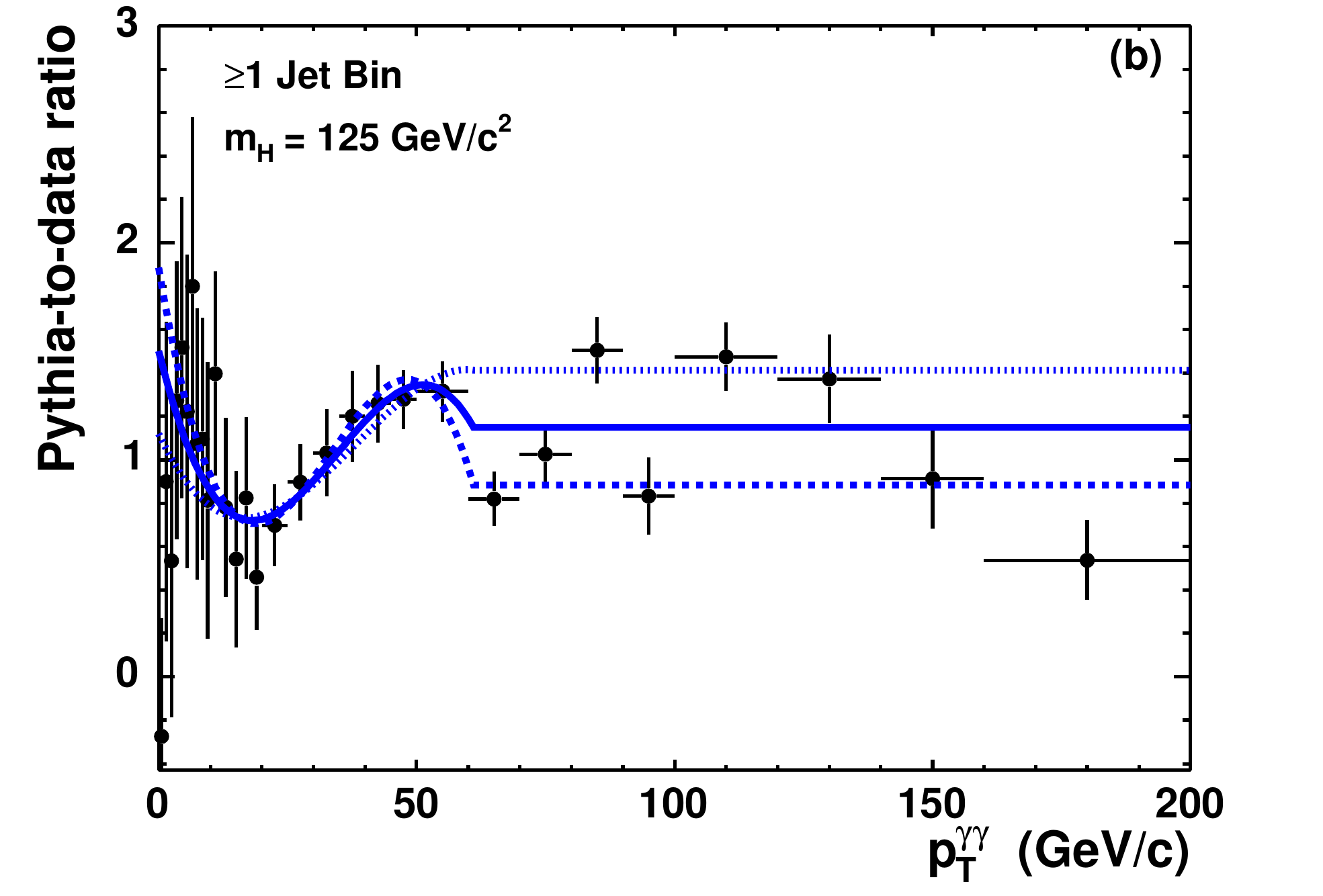}
\caption{For a Higgs boson mass of 125~\gevcc, the reweighting function obtained from the ratio of 
the $p_{T}^{\gamma\gamma}$ distribution in \py\ to the $p_{T}^{\gamma\gamma}$ distribution in 
prompt diphoton data, for events with (a) zero jets and (b) at least one jet. In both plots, the best fit to  
the \py-to-data ratio points is given by a solid curve. 
The other two curves show the systematic uncertainty of the fit.
\label{fig:reweight}}
\end{figure}

In the diphoton cross section analysis~\cite{Aaltonen:2011vk} it was found that a 
\ptgg-dependent correction was needed for the \py\ modeling.  We adopt the 
correction for this analysis, reweighting the \ptgg\ distribution from \py\ to match the 
\ptgg\ distribution from control regions in prompt diphoton data. For each category, 
CC0 and CCJ, and for each Higgs boson mass hypothesis, event weights are derived 
based on the sideband regions, excluding the signal mass window. The weights are 
derived by fitting a smooth function to the ratio of the \ptgg\ distribution from the data to 
that from the \py\ prediction. The best fit in the CC0 category is obtained from a 
polynomial (constant) function for \ptgg~$< 50$~\gevc\ (\ptgg~$> 50$~\gevc). A 
different polynomial (constant) function provides the best fit in the CCJ category for 
\ptgg~$< 60$~\gevc\ (\ptgg~$> 60$~\gevc). Figure~\ref{fig:reweight} shows the 
reweighting function for a Higgs boson mass hypothesis of 125~\gevcc. The solid 
curve shows the best fit to the data and the other two curves show the variations 
induced by propagating the 68\% C.L. fit uncertainties to the fitting function. The rise of 
the reweighting function from \ptgg~$\sim 20$~\gevc\ to \ptgg~$\sim 50$~\gevc\ in 
both the CC0 and CCJ categories is interpreted in Ref.~\cite{Aaltonen:2011vk} as an 
effect of parton fragmentation not modeled in \py, which contributes to the prompt 
diphoton production cross section in that range.

The relative contributions of the two background components are obtained from a fit to the 
diphoton data.  Three histograms for each NN input variable are constructed: one from the 
$\gamma\gamma$ background sample after reweighting, one from the \gjjj\ background 
sample, and one from the diphoton data.  Events used for the fit are required to have 
diphoton mass values greater than 70~\gevcc\, and to be outside of the signal mass 
window.  The histograms are then used to build a $\chi^2$ function defined by 
\begin{eqnarray}
\chi^{2} = \sum_\textrm{i=1}^{N_\mathrm{bins}} \ \sum_\textrm{j=1}^{N_\mathrm{variables}} \left [ \frac{(\alpha g_{ij} + \beta f_{ij} - d_{ij})^{2}}{d_{ij}} \right ]
\label{eqn:chi2}
\end{eqnarray}
where $g_{ij}$, $f_{ij}$, and $d_{ij}$ refer to the number of events in the ith bin of the jth 
input variable for the prompt $\gamma\gamma$ background, \gjjj\ background, and 
diphoton data samples, respectively. The sums are over all bins of each input variable for 
which there are at least 5 events in the data, and  the global $\alpha$ and $\beta$ 
coefficients are determined by minimizing the $\chi^{2}$ function. This function is defined 
and minimized separately for each Higgs boson mass hypothesis and for each category 
(CC0 and CCJ). 

A neural network discriminant is trained separately for each mass hypothesis using signal and 
background events. The signal events used in the training are optimized for the SM scenario and 
are composed of GF, VH, and VBF \py\ samples so that the corresponding total numbers are 
proportional to their SM cross section predictions.  The background sample is made by taking a 
portion of the \gjjj\ sample available for each mass hypothesis and adding   $\gamma\gamma$ 
events from \py\ weighted by the ratio $\alpha/\beta$ from the $\chi^2$ fit for the given mass 
hypothesis.

\begin{figure}
\includegraphics[width=8.0cm]{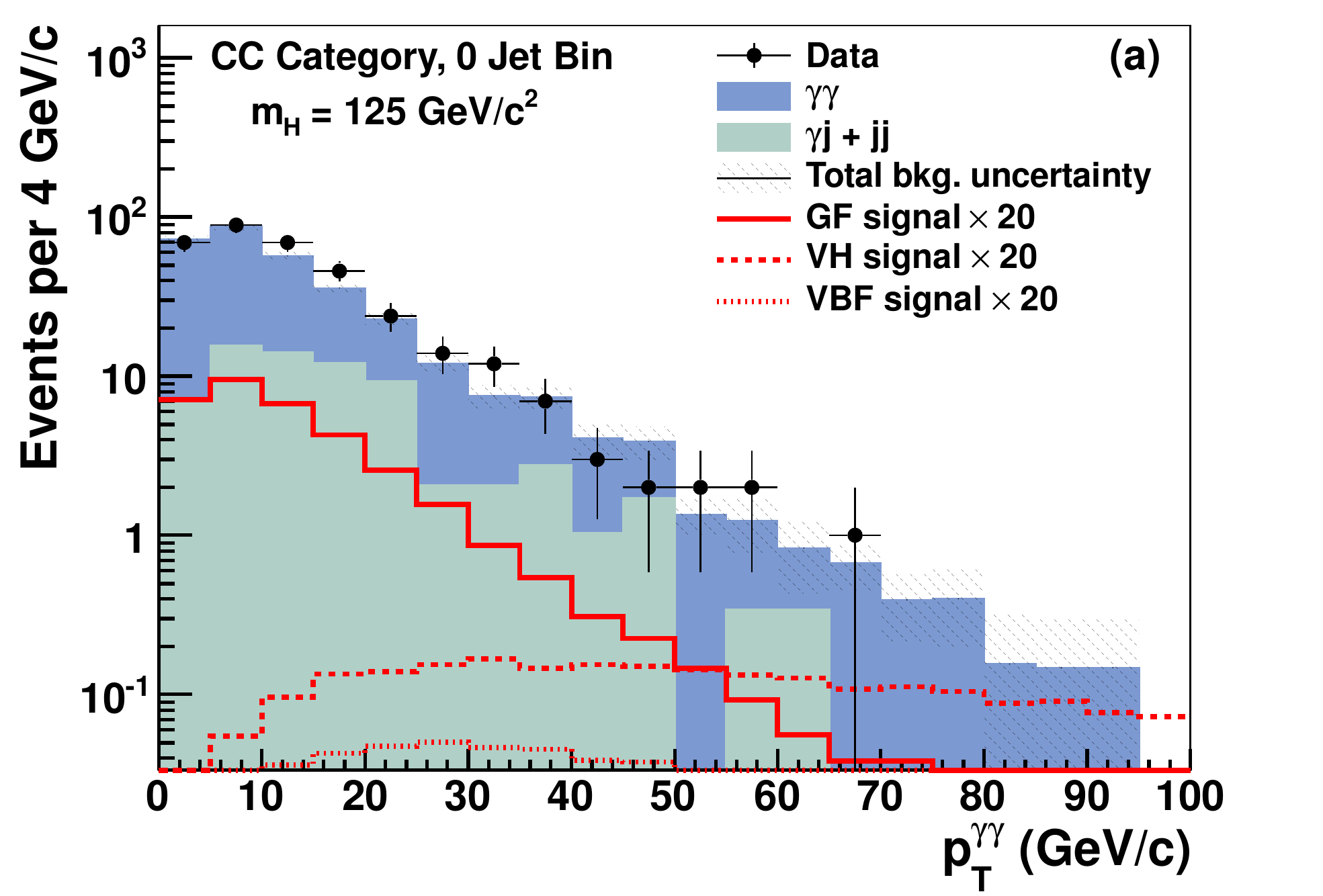}
\includegraphics[width=8.0cm]{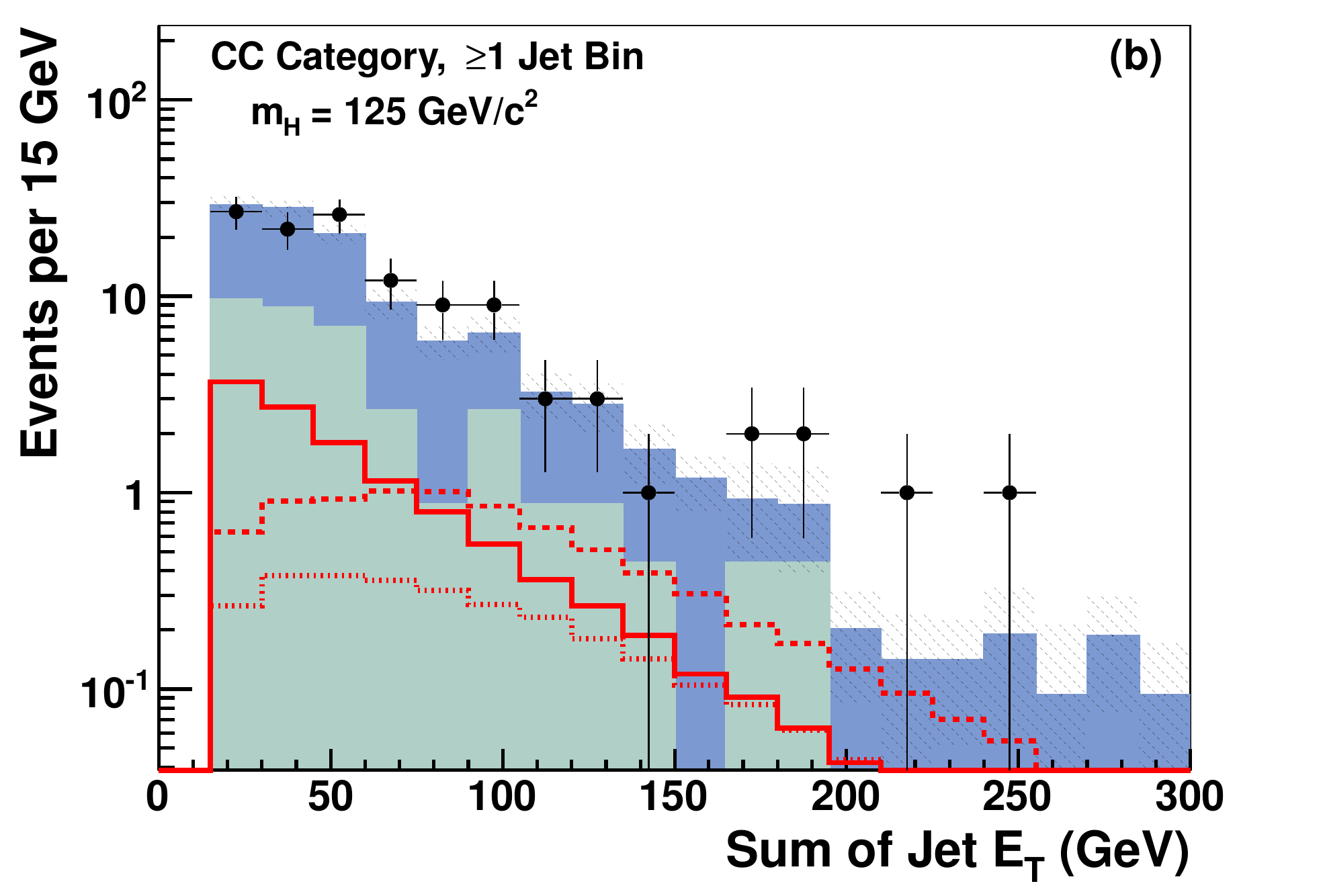}
\caption{For a Higgs boson mass of 125~\gevcc, a comparison of the data to the background 
prediction in (a) the \ptgg\ distribution for the CC0 category and (b) the distribution of the sum 
of the reconstructed jet $E_{T}$ for the CCJ category. The expected SM Higgs boson signal for the 
three production processes is multiplied by a factor of 20.
\label{fig:input}}
\end{figure}

\begin{figure}
\includegraphics[width=8.0cm]{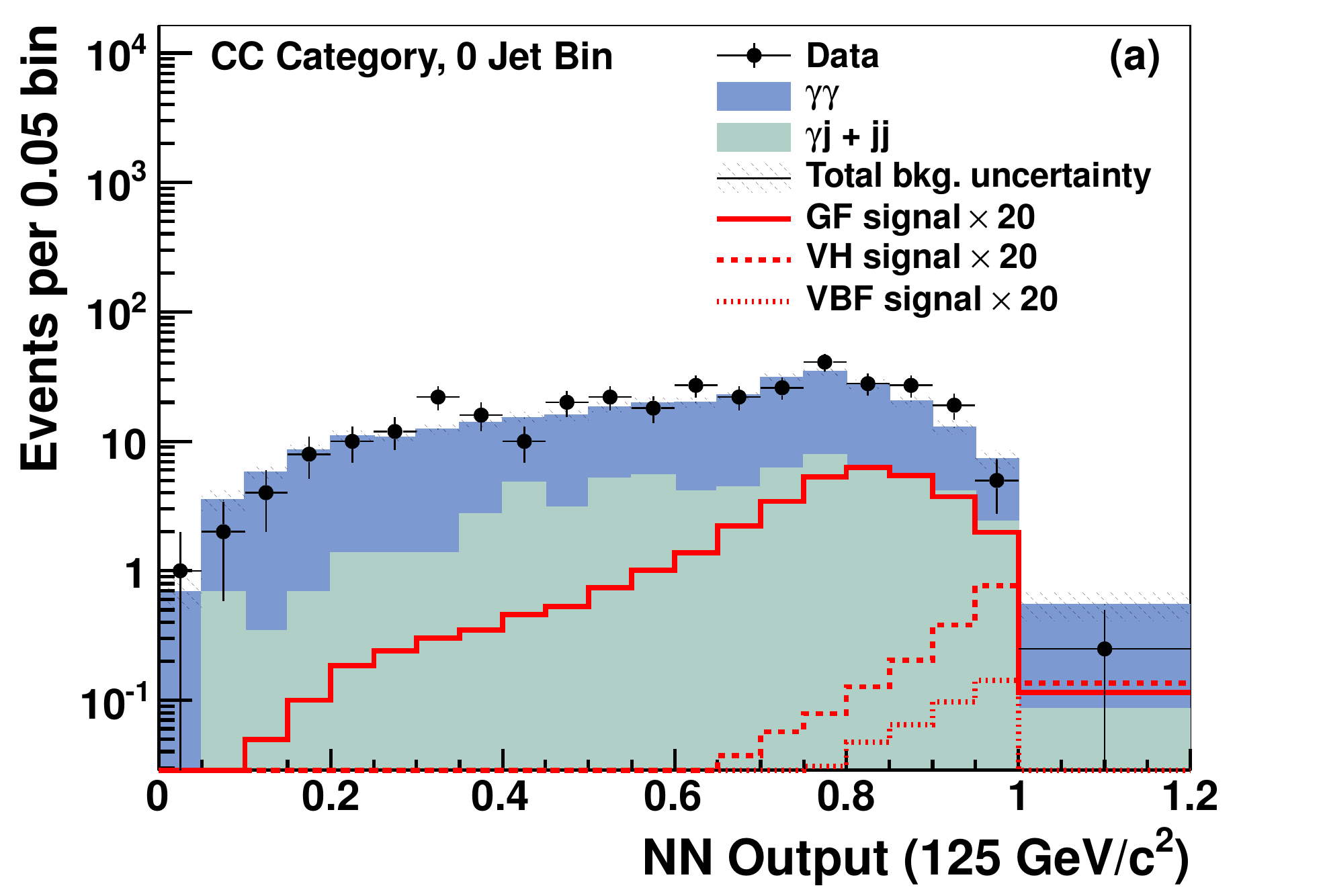}
\includegraphics[width=8.0cm]{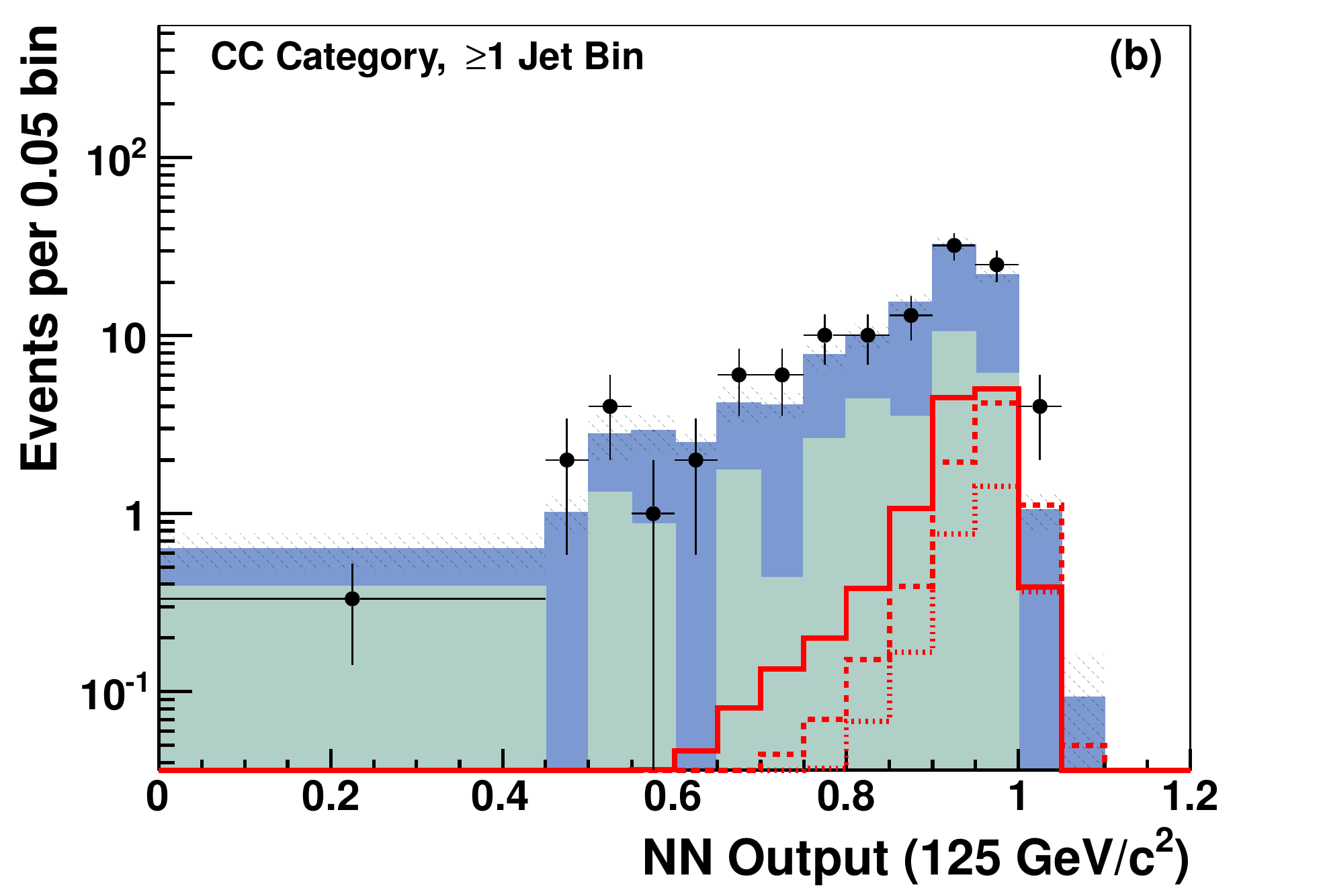}
\caption{For a Higgs boson mass of 125~\gevcc, a comparison of the NN output distributions 
for the data and the background prediction for (a) the CC0 category and (b) the CCJ category.  
The expected SM Higgs boson signal for the three production processes is multiplied by a factor of 20.
\label{fig:NN}}
\end{figure}

After training, the NN is applied to the diphoton data sample. Figure~\ref{fig:input} shows input 
variables such as the \ptgg\ distribution for events with no reconstructed jets and the sum of the 
jet $E_T$ for events with $\geq$1 reconstructed jet. The signal shapes are scaled to 20 times 
the expected number of reconstructed events in the SM scenario. The background prediction 
is also provided. While the $\chi^{2}$ fit described by Eq.~(\ref{eqn:chi2}) is used to fix the 
relative composition of the $\gamma\gamma$ and \gjjj\ background components, the total 
expected number of background events is more accurately determined from sideband mass 
fits, which is the technique described in Section~\ref{sec:Mgg}. The resulting NN shapes for 
\Mh\ = 125~\gevcc\ are provided in Figure~\ref{fig:NN}.

\section{Systematic uncertainties}
\label{systematic}

The sources of systematic uncertainties on the expected number of signal events are the 
same as in the previous CDF \Hgg\ search~\cite{cdfHgg2012}. They arise from the 
conversion ID efficiency (7\%), the integrated luminosity measurement (6\%), varying the 
parton distribution functions used in \py\ (up to 5\%)~\cite{CTEQ6.1M,Weights}, varying the 
parameters that control the amount of initial- and final-state radiation from the parton shower 
model of \py\ (about $4\%$), and the \py\ modeling of the shape of the \ptgg\ distribution for 
the \hf\ signal (up to 4\%)~\cite{MrennaQCD}.  Finally, we include uncertainties from the 
photon ID efficiency (up to 4\%), the trigger efficiency (less than 3\%), and the EM energy 
scale (less than 1\%). 

The statistical uncertainties on the total background in the signal region are 
determined by the fit. They are 4\% or less for the channels associated with the SM 
diphoton resonance search and are less than 7\% for the CC0 and CCJ categories 
used in the multivariate technique. For the channels associated with the 
fermiophobic Higgs boson diphoton resonance search, the background rate 
uncertainty is 12\% or less, except for the high-\ptgg\ bins with conversion photons, 
where it is 20\%. 

For the search using the multivariate technique, in addition to the rate uncertainties 
summarized above, we consider shape uncertainties and bin-by-bin statistical 
uncertainties of the NN discriminant. The signal shape uncertainties are associated 
with initial- and final-state radiation and the jet energy scale~\cite{Bhatti:2005ai}, and 
the background shape uncertainties are associated with the \py\ \ptgg-correction and 
the jet energy scale. The \py\ shape uncertainties due to the \ptgg\ fits are taken as 
uncorrelated between the CC0 and the CCJ categories because the fits determining 
the corrections for each category are done independently. The jet energy scale shape 
uncertainties are correlated between the two categories in order to take into account 
event migration between categories. The dominant uncertainty in the multivariate 
analysis is the bin-by-bin statistical uncertainty of the \gjjj\ background histograms.

\section{Results}
\label{sec:result}

No evidence of a narrow peak or any other structure is visible in the diphoton mass 
spectrum or the NN output distribution.  We calculate a Bayesian C.L. limit for each 
Higgs boson mass hypothesis based on a combination of likelihoods from the 
discriminant distributions for all channels in the corresponding mass signal region. 
The combined limits for the SM search use the NN discriminants of the CC0 and CCJ 
categories and the mass discriminants from the CP, \CCc, and \CPc categories. The 
fermiophobic limits use the NN discriminants of the CC0 and CCJ categories and the 
mass discriminants from the CP, \CCc, and \CPc categories divided into \ptgg\ 
regions. For the limit calculation, we assume a flat prior (truncated at zero) for the 
signal rate and a truncated Gaussian prior for each of the systematic uncertainties.  A 
95\% C.L. limit is determined such that 95\% of the posterior density for $\sigma$~$
\times$~\BRHgg\ falls below the limit~\cite{Beringer2012}.  The expected 95\% C.L. 
limits are calculated assuming no signal, based on expected backgrounds only, as 
the median of 2\thinspace000 simulated experiments. The observed 95\% C.L. limits 
on $\sigma$~$\times$~\BRHgg\  are calculated from the data.  

\begin{figure*}
\includegraphics[width=.49\hsize]{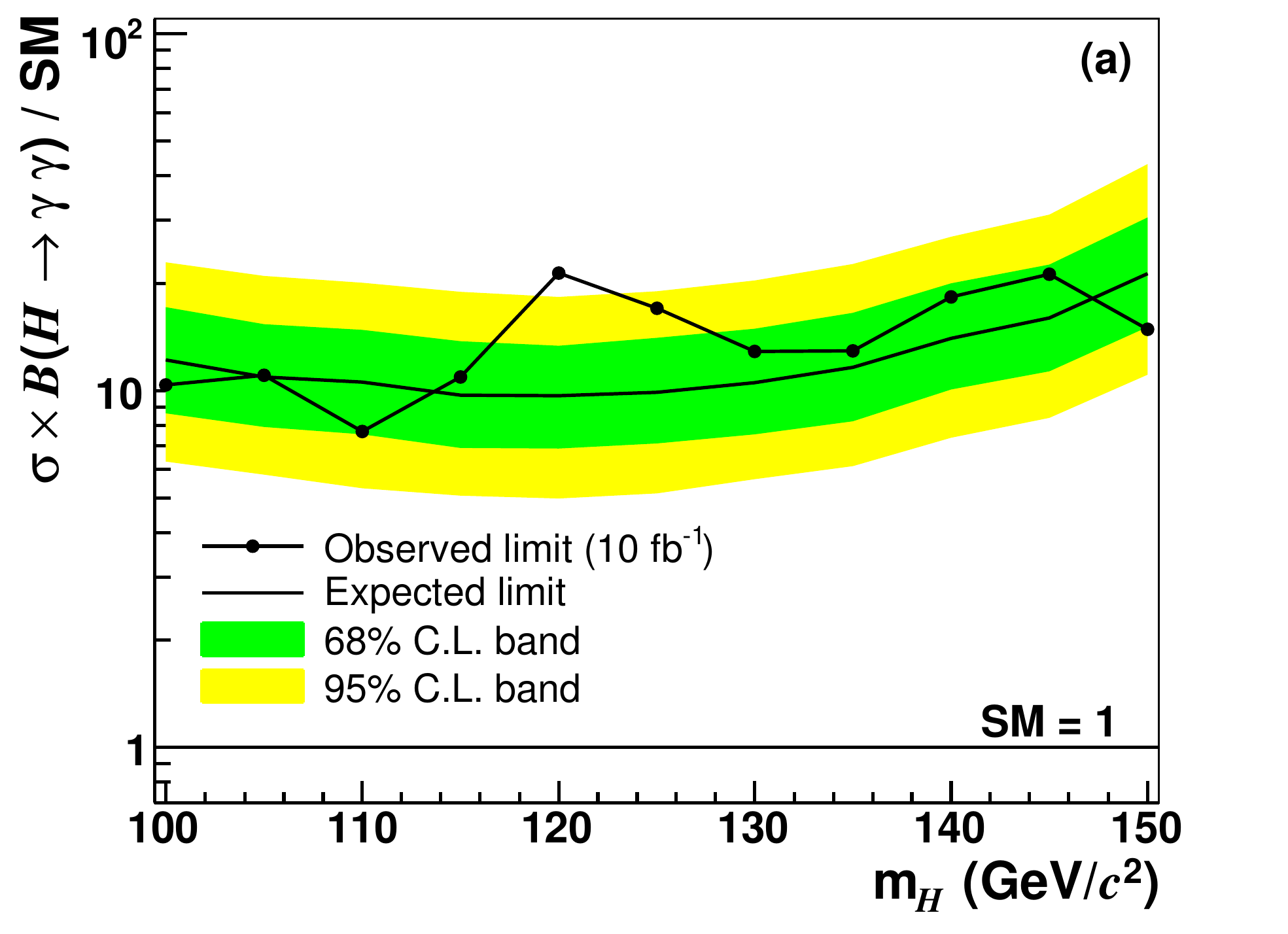} 
\includegraphics[width=.49\hsize]{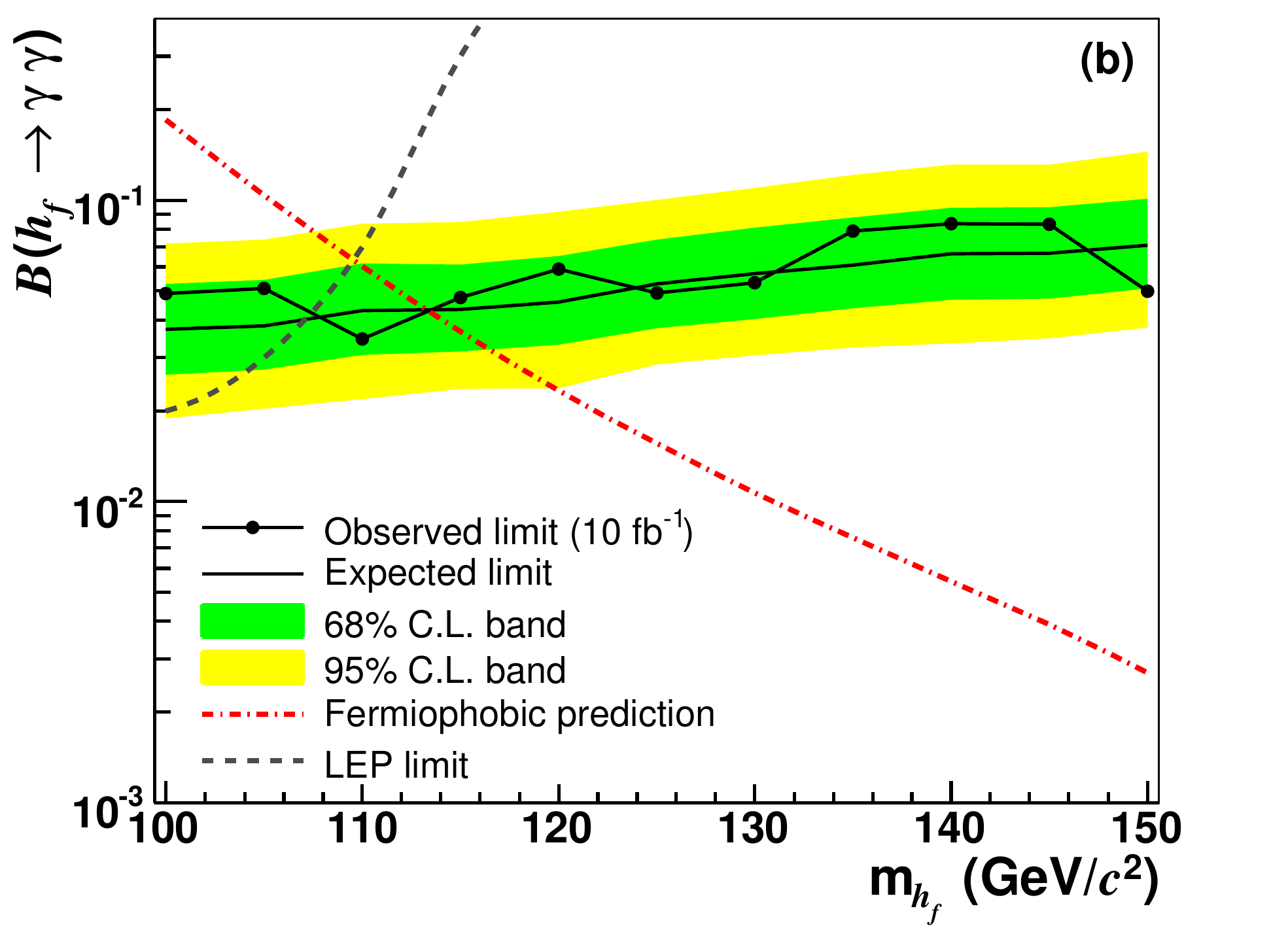} 
\caption{
(a) As a function of \Mh, the 95\% C.L. upper limit on the cross section 
times branching ratio for the SM Higgs boson decay to two photons, 
relative to the SM prediction.  (b) The 95\% C.L. upper limit on the branching 
ratio for the fermiophobic Higgs boson decay to two photons as a function 
of \Mhf.  For reference, the 95\% C.L. limits from LEP are also included.  
The shaded regions represent the 68\% C.L. and 95\% C.L. bands for the 
observed limit with respect to the expected limit based on the distribution 
of simulated experimental outcomes.\label{fig:Limits}}  
\end{figure*}

For the SM Higgs boson search, the results are given relative to the theory 
prediction, where theoretical cross section uncertainties of 14\% on the GF process, 
7\% on the VH process, and 5\% on the VBF process are included in the limit 
calculation~\cite{Botje:2011sn, *Dittmaier:2011ti}. For the \hf\ model, SM cross 
sections and uncertainties are assumed (GF excluded) and used to convert limits on 
$\sigma$~$\times$~\BRHfgg\  into limits on \BRHfgg. The SM and fermiophobic limit 
results for the CC category alone are provided in Table~\ref{tab:CClimits}, showing 
the gain obtained by incorporating a multivariate technique for this category. The 
combined limit results for both searches are displayed in Table~\ref{tab:limits} and 
graphically in Fig.~\ref{fig:Limits}.  Limits are also provided on 
$\sigma$~$\times$~\BRHgg\ for the SM search without including theoretical 
cross section uncertainties. For the SM limit at \Mh\ = 120~\gevcc, we observe a 
deviation of greater than 2.5$\sigma$ from the expectation. After accounting for 
the trials factor associated with performing the search at 11 mass points, the 
significance of this discrepancy decreases to less than 2$\sigma$. When the analysis 
is optimized for the fermiophobic benchmark model, no excess is observed. For the 
\hf\ model, Êwe obtain a limit of \Mhf~$<114$~\gevcc\ by linear interpolation between 
the sampled values of \Mhf\ based on the intersection of the observed limit and the 
model prediction.

\begin{table}
\caption{Expected and observed 95\% C.L. upper limits on the production 
cross section multiplied by the \Hgg\ branching ratio relative to the SM 
prediction for the most sensitive category (CC) using the NN discriminant. For 
comparison, values for the CC category are also provided based on the diphoton 
resonance technique, which uses the \mgg\ shape as a discriminant for setting 
limits. The expected and observed 95\% C.L. upper limits on the \hf\ branching ratio (in \%)  
are provided in parentheses, based on both the NN discriminant and diphoton 
resonance technique for the CC category.
\label{tab:CClimits}}
\begin{center}
\begin{tabular}{ccd@{\quad}cd@{\quad}cd@{\quad}cd}
\hline\hline
$m_{H}$  & \multicolumn{4}{c}{NN discriminant} & \multicolumn{4}{c}{\mgg\ discriminant} \\ 
(GeV$/c^{2}$) & \multicolumn{2}{c}{Expected} & \multicolumn{2}{c}{Observed} &  \multicolumn{2}{c}{Expected} & \multicolumn{2}{c}{Observed} \\ \hline
100 & 13.9 &	(4.6) &	10.6 &	(4.7) &	15.1 &	(5.1) &	11.3	&	(3.5) \\
105 & 12.6 &	(4.6) &	13.0 &	(6.1) &	14.1	&	(5.5) &	10.6	&	(5.1) \\
110 & 11.9 &	(5.2) &	11.8 &	(5.5) &	13.5	&	(5.8) &	11.4	&	(6.3) \\
115 & 11.4 &	(5.2) &	14.1 &	(6.7) &	12.9	&	(6.2) &	15.4	&	(6.0) \\
120 & 11.3 &	(5.5) &	23.2 &	(9.2) &	12.8	&	(6.6) &	22.2	&	(7.3) \\
125 & 11.7 &	(6.4) &	20.5 &	(10.2) &	12.9	&	(6.9) &	21.2	&	(8.0) \\
130 & 12.5 &	(7.0) &	13.1 &	(6.5) &	13.9	&	(7.3) &	16.0	&	(6.0) \\
135 & 13.7 &	(7.7) &	15.0 &	(6.0) &	15.3	&	(7.9) &	17.2	&	(4.9) \\
140 & 16.5 &	(8.2) &	20.4 &	(8.1) &	17.5	&	(8.3) &	25.4	&	(5.9) \\
145 & 18.5 &	(8.4) &	27.4 &	(11.8)&	21.2	&	(8.6) &	24.3	&	(8.8) \\
150 & 25.7 &	(8.7) &	17.1 &	(7.0) &	28.2	&	(9.0) &	15.1	&	(8.4) \\
\hline
\hline
\end{tabular}
\end{center}
\end{table}

\section{Summary and conclusions}
\label{sec:summary}

This Letter presents the results of a search for a narrow resonance in the diphoton 
mass spectrum using data taken by the CDF II detector at the Tevatron. We have 
improved upon the previous CDF analysis by implementing a neural network 
discriminant to increase sensitivity in the most sensitive diphoton category by as 
much as 13\% (17\%) for the SM (fermiophobic) scenario.  In addition, we have included the full CDF diphoton data set, which adds 
more than 40\% additional integrated luminosity relative to the previous diphoton 
Higgs boson search.  There is no significant evidence of a resonance in the data. 
Limits are placed on the production cross section times branching ratio for Higgs 
boson decay into a photon pair and compared to the predictions of the standard 
model and a benchmark fermiophobic model. The latter results in a limit on the 
fermiophobic Higgs boson mass of \Mhf~$< 114$~\gevcc\ at the 95\% C.L. 

\begin{table*}[htp]
\caption{Expected and observed 95\% C.L. upper limits on the production 
cross section times branching ratio relative to the SM prediction, the production 
cross section times branching ratio with theoretical cross section uncertainties 
removed, and the \hf\ branching ratio. The fermiophobic benchmark model 
prediction for \BRHfgg\ is also shown for comparison.
\label{tab:limits}}
\begin{ruledtabular}
\begin{tabular}{ccddddddddddd}
 & \Mh\ (\gevcc) & \mc{1}{c}{100}  & \mc{1}{c}{105} & \mc{1}{c}{110} & \mc{1}{c}{115} &  \mc{1}{c}{120}  &  \mc{1}{c}{125} &
  \mc{1}{c}{130} &  \mc{1}{c}{135} &  \mc{1}{c}{140} &  \mc{1}{c}{145} &  \mc{1}{c}{150}   \\
\hline
\multirow{2}{*}{$\sigma\times$ \BRHgg/SM} & Expected & 12.2& 10.9& 10.6& 9.7& 9.7& 9.9& 10.5& 11.6& 14.0& 16.0& 21.3 \\
& Observed & 10.4& 11.0& 7.7& 10.9& 21.3& 17.0& 12.9& 12.9& 18.3& 21.2& 14.9 \\
\hline
\multirow{2}{*}{$\sigma\times$ \BRHgg\ (fb)} & Expected & 45.1& 39.0& 37.2& 31.8& 29.7& 27.2& 25.5& 24.0& 23.0& 20.4& 20.2 \\
& Observed & 37.9& 40.6& 26.8& 35.9& 66.6& 47.7& 31.5& 26.5& 30.7& 27.2& 13.9 \\
\hline
& Expected                 & 3.7 & 3.8 & 4.3 & 4.3 & 4.6 & 5.3 &	 5.7 & 6.1 & 6.6 & 6.7 & 7.1 \\
\BRHfgg\ (\%) & Observed   & 4.9 & 5.1 & 3.5 & 4.8 & 5.9 & 4.9 & 5.3 & 7.9 & 8.4 & 8.3 & 5.0 \\
& Fermiophobic prediction  & 18.5 & 10.4 & 6.0 & 3.7 & 2.3 &  1.6 & 1.1 & 0.8 & 0.5 & 0.4 & 0.3  \\
\end{tabular}
\end{ruledtabular}
\end{table*}

\begin{acknowledgments}
We thank the Fermilab staff and the technical staffs of the participating institutions 
for their vital contributions. This work was supported by the U.S. Department of 
Energy and National Science Foundation; the Italian Istituto Nazionale di Fisica 
Nucleare; the Ministry of Education, Culture, Sports, Science and Technology of 
Japan; the Natural Sciences and Engineering Research Council of Canada; the 
National Science Council of the Republic of China; the Swiss National Science 
Foundation; the A.P. Sloan Foundation; the Bundesministerium f\"ur Bildung und 
Forschung, Germany; the Korean World Class University Program, the National 
Research Foundation of Korea; the Science and Technology Facilities Council 
and the Royal Society, UK; the Russian Foundation for Basic Research; the 
Ministerio de Ciencia e Innovaci\'{o}n, and Programa Consolider-Ingenio 2010, 
Spain; the Slovak R\&D Agency; the Academy of Finland; and the Australian 
Research Council (ARC). 
\end{acknowledgments}


\begin{thebibliography}{61}
\bibitem{phores}
The energy resolution for photons is 13.5\%/$\sqrt{E_T}$ in the central region and (16\%/$\sqrt{E_T}$) $\oplus$ 1\% in the plug region.
\bibitem{phojet}
Typically, this occurs when a 
jet fragments into a $\pi^0$ or $\eta$ particle that subsequently decays to 
multiple photons, which are then reconstructed as a single photon.
\end{thebibliography}
\end{document}